\documentstyle[12pt]{article}
\begin{document}

\catcode`\@=11
\def\thesection{\arabic{section}}
\def\theequation{\arabic{equation}}


\def\eq#1{\begin{equation}#1\end{equation}}
\def\elab#1#2{\begin{equation}\label{e:#1}#2\end{equation}}
\def\eref#1{Eq.~\ref{e:#1}}
\def\eqsref#1{Eqs.~\ref{e:#1}}
\def\eqarray#1{\begin{eqnarray}#1\end{eqnarray}}
\def\ealab#1#2{\begin{eqnarray}\label{e:#1}#2\end{eqnarray}}
\def\beq{\begin{equation}}
\def\eeq{\end{equation}}

\def\fig#1#2#3{\begin{figure}[htbp]
  \vspace{#3}
  \caption[dummy]{#2}
  \label{f:#1} \end{figure}}
\def\figdbl#1#2#3{\begin{figure*}
  \vspace{#3}
  \caption[dummy]{#2}
  \label{f:#1} \end{figure*}}
\def\tab#1#2#3{\begin{table}[htbp]
  #3
  \caption[dummy]{#2}
  \label{t:#1} \end{table}}

\def\fref#1{Fig.~\ref{f:#1}}
\def\tref#1{Table~\ref{t:#1}}
\def\slab#1{\label{s:#1}}
\def\sref#1{Sec.~\ref{s:#1}}

\catcode`\@=11
\def\rref#1{Ref.~{\csname b@r:#1\endcsname}}
\catcode`@=12


\def\ie{\hbox{\it i.e.}{}}      \def\etc{\hbox{\it etc.}{}}
\def\eg{\hbox{\it e.g.}{}}      \def\cf{\hbox{\it cf.}{}}
\def\etal{\hbox{\it et al.}}	\def\vrs{\hbox{\it vrs.}{}}
\def\dash{\hbox{---}}


\def\deriv#1#2{\frac{d#1}{d#2}}
\def\partder#1#2{{\partial #1\over\partial #2}}
\def\secder#1#2#3{{\partial^2 #1\over\partial #2 \partial #3}}
\def\grad#1{\bigtriangledown #1}
\def\prim#1{#1^{\prime}}
\def\dprim#1{#1^{\prime \prime}}
\def\intd#1{d#1}
\def\pfrac#1#2{ \left( \frac{#1}{#2} \right)}
\def\raisedot{ \raisebox{.25ex}{.}}

\def\simgt{\ \raisebox{-.25ex}{$\stackrel{>}{\scriptstyle \sim}$}\ }
\def\simlt{\ \raisebox{-.25ex}{$\stackrel{<}{\scriptstyle \sim}$}\ }
\def\tr{\mathop{\rm tr}}
\def\Tr{\mathop{\rm Tr}}
\def\Im{\mathop{\rm Im}}
\def\Re{\mathop{\rm Re}}
\def\bR{\mathop{\bf R}{}}
\def\bC{\mathop{\bf C}{}}
\def\bra#1{\left\langle #1\right|}
\def\ket#1{\left| #1\right\rangle}
\def\VEV#1{\left\langle #1\right\rangle}
\def\gdot#1{\rlap{$#1$}/}
\def\abs#1{\left| #1\right|}
\def\contract{\makebox[1.2em][c]{
        \mbox{\rule{.6em}{.01truein}\rule{.01truein}{.6em}}}}


\def\gamfive{\gamma_5}
\def\gam#1{\gamma^#1}


\def\Htwo{{}^2{\rm H}}
\def\Hthree{{}^3{\rm H}}
\def\Hefour{{}^4{\rm He}}
\def\Hethree{{}^3{\rm He}}
\def\Lisix{{}^6{\rm Li}}
\def\Liseven{{}^7{\rm Li}}
\def\Osixteen{{}^{16}{\rm O}}
\def\Ofifteen{{}^{15}{\rm O}}
\def\Nfifteen{{}^{15}{\rm N}}
\def\anti#1{\overline#1}
\def\susy#1{\widetilde#1}
\def\pbar{\anti{p}}
\def\nbar{\anti{n}}
\def\positron{e^+}
\def\antinue{\anti{\nu_e}}
\def\piz{\pi^\circ}
\def\klong{K_L^\circ}
\def\nuebar{\anti{\nu}_e}
\def\photino{\susy{\gamma}}
\def\higgsino{\susy{h}}
\def\zino{\susy{Z}}
\def\gluino{\susy{g}}
\def\wino{\susy{W}}

\def\terra{_\oplus}
\def\solar{_\odot}
\def\Rsol{R\solar}
\def\rau{r\terra}


\def\DM{Dark Matter}
\def\wimp{\chi}
\def\OB{\Omega_B}
\def\OW{\Omega_\wimp}
\def\mW{m_\wimp}
\def\sigva{\VEV{\sigma v}_a}
\def\sigel{\sigma_{\wimp N}}
\def\sigfid{\sigma_{\wimp N38}}
\def\rhofid{\rho_{.3}}
\def\vfid{v_{300}}


\def\en#1#2{{#1} \times 10^{#2}}
\def\M#1{^{-#1}}
\def\UN#1{\ {\rm #1}}
\def\TeV{\UN{TeV}}
\def\GeV{\UN{GeV}}
\def\MeV{\UN{MeV}}
\def\eV{\UN{eV}}
\def\keV{\UN{keV}}
\def\Gauss{\UN{G}}
\def\cm{\UN{cm}}
\def\mm{\UN{mm}}
\def\km{\UN{km}}
\def\sec{\UN{sec}}
\def\yr{\UN{yr}}
\def\Gyr{\UN{Gyr}}
\def\kpc{\UN{kpc}}
\def\Mpc{\UN{Mpc}}
\def\sr{\UN{sr}}
\def\kt{\UN{kt}}
\def\gm{\UN{gm}}
\def\mrad{\UN{mrad}}
\def\mb{\UN{mb}}
\def\erg{\UN{erg}}

\def\degK{\ ^\circ \rm K}
\def\deg{^\circ}
\def\AU{\ {\rm AU}}
\def\flux{\cm^{-2} \sec^{-1}}



\def\phia{\phi_a}
\def\phib{\phi_b}
\def\va{v_a}
\def\vb{v_b}
\def\qa{q_a}
\def\qb{q_b}
\def\na{n_a}
\def\nb{n_b}
\def\wa{w_a}
\def\wb{w_b}
\def\nnu{n_1}
\def\nx{n_2}
\def\phinu{\phi_1}
\def\phix{\phi_2}
\def\chinu{\chi_1}
\def\chix{\chi_2}
\def\vnu{v_1}
\def\vx{v_2}
\def\qnu{q_1}
\def\qx{q_2}
\def\veff{v_{eff}}
\def\mpl{M_{pl}}
\def\lambdaqcd{\Lambda_{QCD}}
\def\Td{T_d}
\def\Tm{T_m}
\def\Tj{T_\chi}
\def\Tid{T_{dec}}
\def\Rj{R_\chi}
\def\Tjo{T_{\chi,0}}
\def\mj{m_\chi}
\def\omegaj{\Omega_\chi}
\def\omegath{\Omega_{\chi,th}}
\def\omegawall{\Omega_{wall}}
\def\gamj{\Gamma_\chi}
\def\tauj{\tau_\chi}
\def\nth{n_{th}}
\def\rhoth{\rho_{th}}
\def\rhomis{\rho_{mis}}
\def\rhocrit{\rho_{crit}}
\def\Tbbn{T_{BBN}}
\def\Rn{R_{n_\chi}}
\def\deltaN{\Delta N_\nu^{BBN}}
\def\ggrav{g_{grav}}
\def\N{N}
\def\Np{N'}
\def\nup{\nu'}
\def\half{\frac{1}{2}}
\def\vold{v}
\def\vh{v_H}

%
\def\d{\delta}
\def\g{\gamma}
\def\k{\kappa}
\def\G{\Gamma}
\def\endtitle{\par\end{quotation}\vskip3.5in minus2.3in\newpage}
\def\NP{Nucl. Phys. B}
\def\PL{Phys. Lett. }
\def\PR{Phys. Rev. Lett. }
\def\PRD{Phys. Rev. D}



 {\hbox to\hsize{Dec. 1992 \hfill UM-TH-92-31}}
 {\hbox to\hsize{\hfill Bartol 92-77}}\par

\begin{center}
\vglue .06in
{\Large \bf {Planck Scale Symmetry Breaking\\ and Majoron Physics}}\\[.5in]

{\bf I.Z.~Rothstein$^1$, K.S.~Babu$^2$ and D.~Seckel$^2$}
\\[.05in]
{\it 1) Department of Physics\\
University of Michigan, Ann Arbor, MI 48109}\\[.15in]

{\it 2) Bartol Research Institute\\
University of Delaware, Newark, DE 19716 }\\[.15in]

{Abstract}\\[-.1in]
\end{center}
\begin{quotation}
Majoron models provide neutrino masses via the spontaneous breaking of
a global $U(1)$ symmetry. However, it may be argued that all global
symmetries will be explicitly violated by gravitational effects. We show
that it is possible to preserve most of the usual features of majoron models
by invoking $U(1)_{B-L}$ to be a gauge symmetry and adding a second singlet
scalar field. The majoron gets a small model dependent mass.
The couplings of majorons to neutrinos may be of ordinary strength
or may be made arbitrarily weak. We discuss the cosmological and
astrophysical consequences of majoron models in the context of a model
dependent majoron mass and neutrino coupling. For an
appropriate choice of parameters majorons can play the role of dark
matter.
\endtitle

\section{Introduction}

Majoron models~\cite{cmp,GelminiR} were designed to accommodate
a wide range of neutrino masses. In most theories, small neutrino masses
are achieved via the see-saw mechanism~\cite{seesaw}.
In grand unified models, seesaw neutrino masses are
naturally of order $10^{-2}$\eV\ or less, but in
majoron models the scale of lepton number violation is left as a free
parameter, and so larger neutrino masses are not implausible. Another
attractive feature of majoron models is that they contain a spontaneously
broken global symmetry, complete with a Nambu-Goldstone boson, \ie,
the majoron. This allows for neutrinos with mass greater
than 30\eV\ to escape the constraint from the closure density of the
Universe~\cite{kt}, since they may decay into a lighter neutrino and a majoron.

Besides allowing for neutrino mass and neutrino decay, majoron models
lead to a host of other cosmological, astrophysical, and
laboratory consequences~\cite{mhptra}.
Since the majoron couples to neutrinos there exist a number of constraints
based on possible majoron production in neutrino rich
environments; \eg\ excess cooling of supernova cores by majoron
emission~\cite{choi} or, a majoron contribution
to the energy density of the Universe which could destroy the
successful predictions of big-bang
nucleosynthesis~\cite{BertoliniS}.
Couplings to other standard model particles arise through loops, which
leads to constraints based on the evolution of red giants~\cite{redgiant} and
horizontal branch stars. Majoron models also possess global $U(1)$ cosmic
strings, although we know of no important consequences of such relics.
Finally, there are a number of laboratory experiments which are sensitive
to majoron models. The most important examples are the width of the $Z$
boson as measured at LEP~\cite{LEP},
which eliminates the simplest of triplet majoron models~\cite{GelminiR};
and the absence of 3-body neutrinoless double beta decay
events~\cite{doublebeta}. We mention these features now because
they illustrate the richness of
majoron models, and they are all modified by our attempt
to reconcile majoron models with short distance symmetry breaking induced
by gravitational effects.

A criticism of majoron models is that the global
symmetry is not necessitated by the known particle spectrum. Although there
may be nothing wrong with this, it has been argued that quantum gravitational
effects should explicitly violate all global symmetries~\cite{Preskill}.
Assuming this to be true, a simple way to save
the majoron models is to ensure that the global symmetry
be {\it accidental}.
This can be accomplished if we gauge $U(1)_{B-L}$ and choose the
quantum numbers of the scalar fields to ensure that
there are no gauge invariant operators
of dimension 4 or less which break the accidental symmetry.
The higher dimensional operators which explicitly break the
accidental $U(1)$ will be suppressed by powers of $\mpl$.

A small explicit symmetry breaking has two immediate
consequences. First, the majoron will acquire a small mass, which in turn
has many astrophysical and cosmological implications. Second, in a
cosmological setting, a U(1) symmetry that is broken both spontaneously and
explicitly implies the formation of a network of strings and walls.
The ability of this network to dissipate places constraints on
the details of the model.

In the usual majoron models there is a direct relationship between
neutrino mass and the coupling of neutrinos to majorons. In the present
context there must be two scalar fields in order to break both
the gauge symmetry and the accidental global
symmetry. Both must have charge under $B-L$, but only one may
couple to right-handed neutrinos to power the see-saw mechanism. The
majoron is predominantly the scalar with the smaller vacuum expectation
value (VEV). If the
smaller VEV is also associated with neutrino masses, then the linear
relationship between neutrino mass and coupling
to majorons is preserved; but if the
larger VEV is associated with neutrino mass the coupling to majorons
will be suppressed by a ratio of small to large VEVs.

This paper gives a more in depth discussion of the issues just summarized.
Section 2 reviews the singlet majoron model and its phenomenology.
Section 3 discusses one way to build majoron models containing an
automatic symmetry to protect against gravity.
Section 4 discusses the cosmological constraints that arise from giving
the majoron a mass. The possibility that majorons play the role
of dark matter is also examined in this section.
Section 5 is a discussion of the model constraints that arise from
a consideration of the string-wall network. Sections 6 and 7
review the constraints from big bang nucleosynthesis and SN~1987A
respectively. The last section contains a brief summary.

Similar cosmological and astrophysical analyses
have been performed for axion models with
gravitational effects~\cite{axiongrav,BarrS}, and for a broad
class of models where the mass of the Goldstone boson is due to instanton
effects in an unspecified strongly interacting sector of the
theory~\cite{FriemanJ}.
The situation for axion models differs from the majoron models presented
here due to upper limits on the neutron electric
dipole moment, which place severe constraints on the strength of explicit
symmetry breaking due to gravitational effects.
These translate into
severe constraints on the model, \eg\ the lowest dimensional operator that
is allowed from strong CP considerations is $d = 10$. No similar
constraint exists for majoron models, and so $d = 5$ is viable.

The phenomenology of textures~\cite{textures}
and neutrino masses~\cite{AkhmedovBS} have also been reexamined in the
context of planck suppressed symmetry breaking.
While this work was in preparation we became aware of a paper
by Akhmedov \etal~\cite{AkhmedovBMS} which considers some of the
same issues as the present paper for majoron physics.

\section{The Minimal Majoron Model}

The simplest majoron model~\cite{cmp} is built by adding  $SU(2)_L\times
U(1)_Y$ singlet neutrinos ($\Np$)
\footnote{We use primes to denote the interaction
basis for the neutrinos and leave off the primes to denote the mass
basis.}
as well as a singlet scalar $(\phi)$ to the spectrum of the standard
model. The singlet scalar couples only to the singlet neutrinos because
it carries lepton number $L=-2$. The relevant part of the Lagrangian
which carries these extra fields is given by
\ealab{lmajold}{
L_{new} & = &  \half \bar{\Np_i} \, i \! \! \not\!\partial \Np_i
   + V(H,\phi) \nonumber \\
& &+ \left\{ \frac{1}{4} \lambda_{ij}^{(1)} \bar{\Np_i} (1 + \gamfive) \Np_j
 \phi
  + \frac{1}{4} \lambda_{ij}^{(2)} \bar{\nup_i} (1+ \gamfive) \Np_j H
           + h.c. \right\}
 ~,
}
where $H$ denotes the standard model Higgs doublet and $i,j$ are
generation indices.
We use four component majorona spinors ($N^c = N,~\nu^c = \nu$),
and assume, for simplicity,
that the Yukawa coupling matrices $\lambda^{(1)},\lambda^{(2)}$ contain no
CP violating phases. The Higgs potential can be written as
\elab{phipot}{
 V = \k_1 [HH^{\dagger}-\vh^2]^2
 + \k_2[HH^{\dagger}-\vh^2][\phi \phi^* - \vold^2]
 + \k_3[\phi \phi^* - \vold^2]^2~.
}
If  $4\k_1 \k_3 >\k_2^2$, then $\VEV{H} = \vh$  and
$\VEV{\phi} = \vold$.
Utilizing the non-linear realization of the global symmetry, the scalar
may be written as

\elab{phidef}{
\phi = (\rho+\vold) \exp(i\chi / \vold),
}
where the majoron will be $\chi$, and the real scalar piece $\rho$ will
be massive.      
We can eliminate the majoron field from the mass matrix by transforming
the neutrino fields;
$\Np_i \rightarrow \exp(-i \chi \gamfive/2v) \Np_i$ and
$\nup_i \rightarrow \exp(i \chi \gamfive/2v) \nup_i$;
leaving
\ealab{lmajolda}{
L  & = & \frac{1}{2} M_{ij} \bar{\Np_i} \Np_j +
 \frac{1}{2} m_{D,ij} \bar{\nup_i} \Np_j +
 \half \bar{\Np_i} \, i \! \! \not\!\partial \Np_i
    \nonumber \\
 & & \mbox{} + \partial_\mu \chi(\bar{\nup_i} \gamma^\mu \gamma_5 \nup_i -
   \bar{\Np_i} \gamma^\mu \gamma_5 \Np_i)/(4 \vold),
}
where
\elab{mnu}{
m_{D,ij} = \lambda_{ij}^{(2)} \vh,~
  M_{ij} = \lambda_{ij}^{(1)} \vold,
}
the subscript $D$ indicates a Dirac mass term coupling $\nup$ and
$\Np$, and we have not written the coupling of neutrinos to $\rho$
particles since we are not interested in theories where the $\rho$ is
light. This will yield the following see-saw mass matrix
\elab{seesaw}{
 L_{mass} = 1/2 \left(
   \begin{array}{cc} \bar{\nup} & \bar{\Np} \end{array}\right)
 M \left( \begin{array}{c} \nup \\ \Np \end{array}\right)
}
where
\elab{massmat}{
 M = \left( \begin{array}{cc}  0 & m_{D,ij} \\
                        m_{D,ij}^T & M_{ij}  \end{array}\right).
}
Upon diagonalization
this mass matrix gives the usual see-saw spectrum of three light
neutrinos ($\nu$) with masses $m_\nu \sim (m_D^2/M$) and
three heavy neutrinos $(\N)$ with masses $O(M)$.

It was pointed out long ago~\cite{schec}
that the mass matrix and the majoron coupling matrix for the
light neutrinos will commute up to $O((m_D/M)^4))$. Therefore, unless we
make the heavy scale $\vold < O(100\GeV)$ the lifetime of any neutrino
$\nu$ with mass greater than 30\eV\ will be too small to avoid the
constraint from the age of the universe, not to mention the more severe
constraint on the lifetime from structure formation~\cite{SteigmanT}.
As a result, it has been generally accepted that additional
Higgs fields must be added in order to get viable models for
neutrino decay~\cite{morehiggs}.
However, it has been recently pointed out~\cite{palash} that the result
in reference~\cite{schec}
does not hold in general. Radiative corrections will generate finite pieces
of $O(m_D^2/(8\pi^2 M^2))$ to the off diagonal part of the light
neutrino couplings which {\it will not} be the same as the
corrections to the off-diagonal terms of the mass
matrix\footnote {It is also possible
to get rapid decay if one chooses the global symmetry
to be a linear combination of the lepton
numbers ($L_{e,\mu,\tau})$ as opposed to just $L$~\cite{babu}.
Rapid decay into three neutrinos is possible in models with Dirac neutrino
masses and extra singlet scalars which do not acquire VEVs~\cite{3nu}.}.
The light neutrino Lagrangian may then be written with derivative
couplings as
\ealab{nunuj}{
L_\nu = & & \half m_{\nu,i} \bar{\nu_i} \nu_i \nonumber \\
 & + & \frac{\partial_\mu \chi}{4\vold}
  \left(\bar{\nu_i} \gamma^\mu \gamma_5 \nu_i +
 \sum_{ij} \bar{\nu_i} \gamma^\mu
   (g_{v,ij} + g_{a,ij} \gamma_5)\nu_j \right),
}
where $g_v$ and $g_a$ are of order $O(m^2/(8\pi^2 M^2))$ and there
may be vector off-diagonal couplings, \ie\ $g_{v,ii} = 0$.
The decay rate for the heaviest of the light neutrinos into a particular
channel is then given by
\elab{nudec}{
\G = \frac{m_\nu}{64\pi} \pfrac{m_\nu}{\vold}^2
   (g_{v,ij}^2 + g_{a,ij}^2),
}
where we have neglected the mass of the final state neutrino.

In the singlet majoron model, coupling to other fermions arises
through an intermediate $Z$ boson and a neutrino loop, see \fref{eechi}.
The electron coupling in pseudoscalar form is then
\elab{eej}{
g_{ee\chi} \approx \frac{G_F m_e m_{\nu}}{16\pi^2}.
}
The most important constraint on this coupling arises from consideration
of the cooling of red giant stars~\cite{redgiant},
$g_{ee\chi} >10^{-13}$.

\section{The Singlet Majoron Model with Explicit Symmetry Breaking}

Assuming that short distance physics only respects local gauge symmetries,
the only way to have Nambu-Goldstone bosons appear in the spectrum is to ensure
that there is an $accidental$ (clearly a misnomer in this case) symmetry.
There is no way to prevent operators which break
the accidental symmetry from appearing at some level, but we can
arrange for these operators to have dimension greater than four. In
this case the symmetry breaking operators will be suppressed by
powers of $\mpl$ and a small mass is induced for the Nambu-Goldstone boson.

To accomplish this for majoron models, we gauge $U(1)_{B-L}$ and include in
the spectrum two singlet scalar fields. The first, $\phinu$, plays a
similar role to $\phi$ in the singlet model of section 2. This
field has $B-L$ quantum number $\qnu = 2$,
allowing it to have a Yukawa coupling with the singlet
neutrinos. The $U(1)_{B-L}$ charge of the additional scalar field,
$\phix$, is chosen such that the lowest dimensional $B-L$
invariant operator which breaks the accidental $U(1)$ has dimension
greater than four. To leave matters as general as possible we allow
$\phix$ to have fractional $B-L$ charge. As such,
there will be several different models possible for a given dimension;
for example, $\qx = 1/2$, 4/3, 3, or 8, will result
in a dimension 5 operator suppressed by one power of $\mpl$. As examples
of $\qx$ assignments that do not work, $\qx = 1$
or 4 allow a planck scale cubic term in the lagrangian and the global
symmetry would disappear entirely from the low energy lagrangian.

We write the lowest dimensional symmetry breaking term as
\elab{operator}{
L_{S.B} = \ggrav \phinu^{\nnu} \phix^{*\nx} / M_{pl}^{d-4} + h.c.,
}
where $\ggrav$ is the dimensionless strength of the operator,
$d = \nnu + \nx$ is the dimension of the symmetry breaking operator,
and $\nnu$ and $\nx$ are integers. Gauge invariance imposes the constraint
\elab{charges}{
\nnu \qnu - \nx \qx = 0,
}
and to ensure that $d>4$ we require
\elab{chargecond}{
 a \qnu - b \qx \neq 0 ~:~ if ~a+b\leq 4;~~a,b \in integers.
}

By choosing an appropriate Higgs potential both singlets will receive
vacuum expectation values, $\vnu$ and $\vx$ respectively.
Given that $B-L$ is a local gauge symmetry there will be a lower bound
on the $Z^\prime$ mass of $O(300 \GeV)$ which, for reasonable $U(1)_{B-L}$
gauge couplings, will also be a lower bound on the larger of the
two singlet VEVs.
Had we chosen to gauge an anomaly free combination of lepton numbers
this constraint would be relaxed~\cite{ray}.

Upon symmetry breaking, one combination of the phases of the singlets
becomes the longitudinal part of the $Z^\prime$ gauge boson,
while the orthogonal combination becomes the majoron,
\elab{majoron}{
\chi = \frac{\qx \vx \chinu - \qnu \vnu \chix}
  {(\qx^2 \vx^2 + \qnu^2 \vnu^2)^{1/2}} .
}
The diagonal majoron couplings to the light neutrinos are given in the
pseudovector form as
\elab{nunujnew}{
L_{\nu\nu\chi} = \frac{\qx \veff}{\vnu}
\frac{\partial_\mu \chi}{4\vnu} (\bar{\nu_i} \gamma^\mu \gamma_5 \nu_i)~,
}
where the first factor accounts for the mixing of $\phinu$ and $\phix$ in
\eref{majoron}, and for convenience, we define
\elab{veff}{
 \veff^2 = \frac{\vnu^2 \vx^2}{\qnu^2 \vnu^2 + \qx^2 \vx^2}.
}
The explicit symmetry breaking will induce a mass for the majoron,
\elab{mj}{
\mj^2 = 2 \ggrav \frac{\vnu^{\nnu} \vx^{\nx}}{\mpl^{d-4}}
  \frac{\nx \nnu}{\qx \qnu \veff^2}.
}

The coupling of other fermions to the majoron arise from the same diagram
as in the simple model of section 2.  At first it might appear
that one could replace the $Z$ boson by a $Z^{\prime}$ boson, while at
the same time replacing the light neutrino loop by a loop with a heavy
singlet neutrino.  If such diagrams existed they would dominate over the
$Z$ exchange by a ratio of heavy to light neutrino masses, however, one
can show that such contributions vanish.  First, the
$\chi -Z^{\prime}$ mixing (through a heavy singlet neutrino loop)
cannot lead to $\overline{f} \gamma_5 f \chi$ vertex since $B-L$ is a
vectorial symmetry.  There will be $Z-Z^{\prime}$ mixing at one--loop via
the exchange of charged fermions and, since $Z$ coupling to matter is
parity violating, $\overline{f} \gamma_5 f \chi$ may be induced as a
two--loop effect.   However, such diagrams also give vanishing
contribution.  This is best seen by working in Landau gauge, where the
gauge boson propagator is purely transverse.  There is no
$\chi-Z^{\prime}$ or $\chi-Z$ mixing in this gauge, so only the unphysical
Higgs may contribute; but this scalar is necessarily an $SU(2)_L$--singlet
(since it has to couple to the heavy singlet neutrino) and so has no
coupling to the standard model fermions.

The constraint from coupling majorons to ordinary matter is therefore the
same as in the simplest model, with the exception that even this mild
constraint does not hold if $\mj \simgt 10\keV$, since the majorons
will be too heavy to be produced in red-giant cores.

\section{Majoron contributions to the energy density}

In theories where a global symmetry is broken spontaneously as
well as explicitly by a term with a relatively small coupling,
there are two
contributions to the energy density in Goldstone bosons, the thermal
production due to interactions with the other particles in the medium and
a coherent oscillation of the Goldstone field if
the initial angle of that field does not align with the direction of
explicit symmetry breaking. The coherent contribution may be
enhanced by the presence of strings and walls associated with
spontaneous symmetry breaking, or may be suppressed by an inflationary
epoch. Both terms may be alleviated if the Goldstone bosons decay rapidly.

We begin by summarizing the cosmological history of the model.
For this task it is convenient to refer to the scalar fields
by the magnitude of their VEVs, in which case we will refer to the
smaller VEV (and its field) by the subscript $_a$ and the larger VEV by
the subscript $_b$. It may be that $a \equiv 1$, indicating the scalar
field that couples to neutrinos, or that $a \equiv 2$,
depending upon the scenario.
Then, the larger of the VEVs, $\vb$, is established first,
the associated scalar is ``eaten" by the $U(1)_{B-L}$
gauge field, and the $Z'$ gets a mass. At this time a set of local $U(1)$
gauge strings will form.

When the remaining global $U(1)$ symmetry is
broken, $\va$ is established and the majoron becomes a
light degree of freedom. Breaking the global symmetry results in a set of
global cosmic strings. Additionally, the local gauge strings present from the
breaking of $U(1)_{B-L}$ may become catalysis sites for
another set of global strings.

The majoron has a small mass (\eref{mj})
which is present immediately at the breaking of $\phia$.
This mass becomes dynamically important at a time
labeled by the temperature $\Tm$, defined by
\elab{tmass}{
\mj = 3H(\Tm),
}
where $H$ is the expansion rate of the Universe. It is possible that
$\mj > H$ at the time when $\va$ is established, in which case
$\Tm \approx \va$. In either case, two things occur at $\Tm$.
First, one may expect a contribution to the energy density from a
non-thermal population of long-wavelength coherent majorons. Second, the
global cosmic strings become the edges of dynamically important domain
walls, whose thickness is of order $1/\mj$. Depending upon details of the
model these string-wall networks may or may not go away.

At first the majoron field should be thermally coupled
to the plasma through scattering processes, but at some time
(labeled by the temperature $\Td$) the majorons will decouple.
It is possible that decays and inverse decays may
recouple the majorons to the plasma at a later date. The rate of
inverse decays in a relativistic plasma is $\sim \gamj \mj/T$, whereas the
expansion rate is $\sim T^2/ \mpl$. Evidentially, inverse
decays can bring a species back into equilibrium at low temperatures -
depending on the interaction strength. If the majoron-neutrino coupling is
weak then full equilibrium will not be achieved until the majorons are
non-relativistic - at which time their energy density will be dumped into
neutrinos but back reactions will be unimportant. Of course, this may
occur after the present epoch.

\subsection{Thermal contribution to the energy density}

If the majorons are lighter than a few\eV\ their thermal contribution
to the energy density will not be significant, but a larger mass
is certainly possible in the present context.
The energy density in a thermal relic population of majorons is
\elab{thdens}{
\rhoth = \nth \mj = \frac{\zeta(3)}{2 \pi^2} \Tj^3 \mj,
}
where $\Tj$ is the `temperature' of the majorons. Generally, $\Tj$ will
differ from the photon temperature by a factor $\Rj$, in which case the
thermal relic density can be given in terms of the critical density for
closure, $\rhocrit$, as
\elab{omegath}{
\omegath \equiv \frac{\rhoth}{\rhocrit}
  = 19.8 \frac{\mj}{\hbox{\rm keV}} \frac{\Rj^3}{h_0^2},
}
where $h_0$ is the expansion rate in units of $100\km\sec\M1\Mpc\M1$,
and we used a microwave background temperature of $2.74\degK$.
If the Universe has expanded
adiabatically since the majorons decoupled, and light neutrinos are the
only non-photonic species present today then
\elab{gj}{
\Rj^3 = \pfrac{\Tj}{T_\gamma}^3 = \left( \frac{g_\gamma}{g_\gamma + g_e}
          \frac{g_\gamma + g_e + g_\nu}{g_*} \right)
   \approx 5.8\times10^{-2},
}
where $g_\gamma = 2$, $g_e = 7/2$, $g_\nu = 21/4$, and $g_*$ is the total
number of relativistic degrees of freedom other than majorons at the time
that the majorons decouple.

As will be shown below, for most scenarios decoupling may be
expected to occur soon after $\phia$ develops a VEV.
At first the self couplings of the majorons and
$\rho$ particles and their coupling to singlet neutrinos
may be expected to keep the majorons in equilibrium.
If $\va$ and $\vb$ are not too dissimilar
the $B-L$ gauge interactions will couple the singlet
neutrinos, majorons and $\rho$ particles to the ordinary degrees of
freedom in the plasma. However, all these particles, except for the majoron,
presumably have masses of order $\va$ or $\vb$, and will soon freeze out of
the primordial plasma. After they freeze out the only way to avoid decoupling
is through the light neutrinos. The most direct reaction to maintain
equilibrium is $\nu \nu \leftrightarrow \chi \chi$ through t-channel
exchange of a neutrino. The cross section for this process
is given by
\elab{sigmanunu}{
\sigma_{\nu\nu\rightarrow\chi\chi} \simeq \frac{1}{64 \pi}
   \frac{m_\nu^2}{\vnu^4} \pfrac{\qx \veff}{\vnu}^4.
}
This result applies in the limit where the
$\rho$ particles are much heavier than the momentum exchange in the
reaction. The last factor arises from projecting the majoron onto the
scalar field that couples to the  neutrinos. Allowing this
factor to be of order unity, the decoupling point is given by
\elab{Td}{
\frac{\Td}{\vnu} \sim 10^2 \pfrac{1 \eV}{m_\nu}^2 \pfrac{\vnu}{1 \GeV}^3 .
}
Unless $\vnu$ is quite small and the neutrino masses are large the light
neutrinos cannot keep the majorons in equilibrium after the phase
transition. The converse
of this is that the constraint on majoron models from big bang
nucleosynthesis is fairly weak (see Section 6). Thus, except for
exceptional circumstances, it is reasonable to assume that the
majorons decouple soon after the phase transition.

\eref{omegath} is valid only so long as the majorons are stable. However,
if majorons have a mass then they may decay with a rate
\elab{chidecay}{
\gamj = \frac{\mj}{16\pi} \sum_i g_{\nu_i}^2  ,
}
where
\elab{nunuchi}{
g_{\nu_i} = \frac{m_{\nu_i}}{\vnu} \frac{\qx \veff}{\vnu}
}
is the diagonal coupling to neutrino $i$ in its pseudoscalar form.
The sum is over neutrino species lighter than the majoron. Off diagonal
couplings are presumably less important.
The decay rate depends on the light neutrino masses, which arise
through the seesaw mechanism. Phenomenological constraints on neutrino
masses will impose constraints on the majoron decay rate.

Some general remarks. a) If neutrino masses are small enough, then majorons
are stable for all practical purposes. b) If the majorons decay well after
they become non-relativistic, the decay neutrinos will contribute to the
energy density in radiation, and this can produce a conflict between
theories of galaxy formation and limits on the fluctuations in the
temperature of the cosmic background radiation. In the extreme case the
density of decay products can overclose the Universe. c) If
the majorons decay too quickly then inverse decays may recouple the
majoron field to the plasma before nucleosynthesis begins at a time of
about 1 sec, or a temperature $\Tbbn \approx 1\MeV$. In this case
an excess of $\Hefour$ will be produced (see section 6).
d) If majorons are more massive than about 10\MeV\ then
their decay before nucleosynthesis (more specifically - before the
neutrinos decouple) poses no significant
cosmological problems other than a potential increase in entropy and
consequent dilution of baryon number.

Let us investigate the possibility that relic majorons
form warm dark matter.
As an example, consider the case where the two VEVs are comparable,
$\vnu = \vx = v$. We vary $\qx$, the $B-L$ charge of $\phix$,
so as to get the appropriate dimension operator.
To get the majoron to be warm dark matter we want the mass of the
majoron to be of order a\keV. Using \eref{mj} and assuming a
dimensionless strength of order unity for the explicit symmetry breaking,
we find a relation between $v$ and the dimension of the operator,
\elab{mjkev}{
\log_{10}(v/1\GeV) \sim \frac{19 (d-4) - 12}{d-2}.
}
The decay rate will be given by \eref{chidecay}. If we scale to a
neutrino mass of $10 \eV$, keep the majoron mass at a\keV, and drop
the numerical factor $(\qx \veff/ \vnu)^2$ then the majoron lifetime is
\elab{taujkev}{
\tauj \sim 3 \times 10^5 \pfrac{v}{\hbox{\rm TeV}}^2
  \pfrac{\hbox{\rm 10 eV}}{m_\nu}^2 \pfrac{\hbox{\rm keV}}{\mj} \sec.
}
If $d=5$ then $v \sim 200\GeV$ and the lifetime is around $10^4$ seconds.
Majorons have mass but decay quickly so they do not constitute dark
matter. For\keV\ majorons, their decay rate lies in an acceptable window
which poses no problems for galaxy formation or for
nucleosynthesis. If $d = 6$ then $v \sim 3\times10^6 \GeV$ and the
lifetime is around $3\times10^{12}\sec$. This decay rate is troublesome
for galaxy formation, but a modest decrease in the heaviest neutrino mass
would allow the majoron to be sufficiently stable to be dark matter. For
$d = 7$ we find $v = 10^9\GeV$ and lifetimes longer than the age of the
Universe, so majorons can play the role of warm dark matter in such
models. Higher $d$ also admit a value of $\omegaj \sim 1$ in stable
thermal relics. Clearly, we could vary the ratio of the two VEVs, the
dimensionless strength of the gravitational operator, \etc, providing even
more flexibility to the models.

Let us look at the $d=5$ case a little more closely. This can be achieved
by $\qnu = 2$ and $\qx =$ 1/2, 4/3, 3, or 8. Let us concentrate on
$\qx = 1/2$, and still take the two VEVs to be the same.
Assuming that both VEVs develop before the standard model breaks from
$SU(2) \times U(1)_Y$ down to $U(1)_{em}$, so that $g_* = 67\frac{3}{8}$,
we find $\Rj = 0.058$. For a Hubble parameter of $h = 0.5$
(where $h$ is $H$ in units of
100 \km/\sec/\Mpc) and a majoron mass of $\mj = 53\ggrav^{1/2} \keV
(v/1\TeV)^{3/2}$ this gives $\omegaj = 242 \ggrav^{1/2} (v/1\TeV)^{3/2}$, if
the majorons were stable. However, with these parameters and
a 30\eV\ neutrino mass, the lifetime of the majorons is
of order $10^4$ seconds.

Coincidentally, the temperature of the Universe at $10^4$ seconds
is of order 10\keV, so the majorons would be mildly non-relativistic when
they decay. The decay neutrinos would be slightly more energetic than
the blackbody neutrinos, but not so much as to pose a problem for galaxy
formation. However, as Akhmedov, \etal~\cite{AkhmedovBMS} point out,
if one assumes an upper limit on the neutrino mass of roughly $30 \eV$,
then this argument yields an upper limit to the VEV, $v \simlt 10\TeV$.
Both $\mj$ and $\tauj$ increase with $v$ and so the present density in
relativistic decay neutrinos also increases; which eventually creates
a problem for galaxy formation. In that analysis there is only one
VEV, but in our models there is more flexibility. By separating $\vnu$ and
$\vx$ and choosing different $B-L$ charges one can maintain a
short majoron lifetime while increasing $\mj$.
For example, in the toy case we chose $\qnu = 2$ and $\qx = 1/2$ so that
varying $\vnu$ and $\vx$ independently (but keeping $\vnu < \vx$, so that
$\veff \approx \vnu$) gives
\ealab{twovevs}{
   \mj & \approx & \frac{\vnu^{-1/2} \vx^2}{\mpl^{1/2}} \\
   \tauj & \approx & \frac{8 \pi}{\mj} \pfrac{\vnu}{m_\nu}^2
}
As $\vx$ is raised the majoron mass increases but the lifetime shortens
since the neutrino couplings haven't been altered. The energy in
relativistic decay products increases more slowly. There is still a
prohibited region from BBN considerations,
$0.1\MeV \simlt \mj \simlt 10 \MeV$, but generally more
massive majorons are still viable in a model with two VEVs.

Another interesting case occurs when the
heaviest neutrino has a mass chosen to solve the
solar neutrino problem via the MSW mechanism~\cite{msw},
$m_\nu \sim 10^{-2.5} \eV$.
The majoron lifetime would be around $10^{12}$
seconds in such a model. This value is somewhat uncomfortable - the decay
products would be relativistic and the Universe would be dominated by
radiation for a period late in its history, making a consistent theory of
galaxy formation more difficult. To rectify
this, it would be desirable to make the majoron lighter and longer lived than
the age of the Universe.  This could be accomplished by increasing $\vnu$
while holding $\vx$ fixed and the neutrino masses constant.
This would decrease the majoron mass and their coupling to neutrinos.

Finally, one of the main motivations for majoron models is to allow for
neutrino masses in excess of 30\eV. The presence of the massless majoron
allows the heavy neutrino to decay into a lighter neutrino and a
majoron with a short lifetime. The issue of off diagonal couplings may be
addressed by invoking radiative loops, or enlarging the model to include
extra scalar fields so that the mass and coupling matrices are not nearly
proportional. With the consideration of gravitational effects one may
expect the majoron to get a mass, perhaps in the\keV\ region. Neutrinos
heavier than the majoron may still decay into a majoron and a lighter
neutrino, but neutrinos less massive than the majoron, and yet more
massive than 30\eV\ have no obvious decay channel and would present a
cosmological problem.

\subsection{Coherent Oscillations}

The small explicit breaking of a spontaneously broken global symmetry
leads to a coherent oscillation of the Goldstone field around its minimum.
This phenomenon is well known in axion models~\cite{preskill} and leads to a
constraint on the axion decay parameter, $f_a$. If gravitational effects
explicitly break the Peccei-Quinn symmetry in an axion model,
then the energy in coherent oscillations is
smaller than that in thermal axions unless the explicit symmetry breaking
operator has $d \ge 10$~\cite{axiongrav,BarrS}. The models considered here are
similar except for the presence two VEVs, but we will
show that the second VEV can only make the coherent oscillations less
important.

The majoron field is static until the mass becomes greater than the
expansion rate of the Universe, $\mj = 3H(\Tm)$.
The energy density in majorons at that time is
\elab{rhomis}{
 \rhomis = {1\over{2}} \mj^2 \chi^2_0 \frac{\qx \qnu \veff^2}{\nx \nnu},
}
where $\chi_0$ is the initial misalignment angle, presumably of order
unity. These coherent oscillations have a wavelength which is greater
than their Compton wavelength and the energy density scales as
non-relativistic matter. Taking into account the expansion of the universe
from $\Tm$ until
today yields
\elab{omegamis}{
\omegaj \approx 7 \times 10^{-20} \frac{g_m^{3/4}}{h^2}
     \frac{\mj^{1/2} \veff^2}{\hbox{\rm GeV}^{5/2}}
     \frac{\qx \qnu}{\nx \nnu},
}
where $g_m$ is the number of equilibrated degrees of freedom at $\Tm$.

To compare the thermal relic and coherent energy densities we compare
their contributions to the majoron number density at the temperature $\Tm$.
Dropping all dimensionless factors, the ratio is given by
\elab{Rn}{
\Rn \equiv \frac{n_{mis}}{n_{th}} \approx
   \frac{\mj \veff^2}{\mj^{3/2} \mpl^{3/2}}
 \approx  \pfrac{\va}{\mpl}^{(10-d)/4} \pfrac{\va}{\vb}^{\nb/4}
}
where we have used $\Tm \approx (\mj \mpl)^{1/2}$ in the first relation;
and \eref{mj} for the majoron mass and \eref{veff} for $\veff$ in the second.
Note that the result depends only on the relative magnitude of the VEVs
and not on which one couples to neutrinos. The power $(10-d)/4$ is the
same as in the axion result. The extra ratio of VEVs reflects the fact
that a large $\vb$ will increase $\mj$ without affecting $\veff$, but this
can only reduce $\Rn$. Finally, for $d=5$ (and maybe for $d=6$)
$\Tm \approx \veff$, \ie\ the majoron mass is greater than the expansion
rate at the moment of the phase transition. In this case
$\Rn \approx \mj/\veff$, and, again, misalignment is unimportant.
Thus, the misalignment energy is only important if the dimension of
the gravitational operator is $d \ge 10$.

\eref{omegamis} may need to be modified to account for the majorons that
arise from cosmic strings. Following the arguments
made by Davis and Shellard~\cite{DavisS} for axion models,
the presence of global strings may be expected to
increase the number of majorons in the coherent field by perhaps an
order of magnitude due, primarily, to the logarithmic divergence in
the string tension of global strings. However, this conclusion
depends on the spectrum of Goldstone bosons radiated
from cosmic string loops at the time when $T = \Tm$~\cite{HarariS}.
A conclusion to the argument awaits a
definitive numerical analysis of the radiation spectrum.

The derivation of \eref{omegamis} also assumed that the initial
misalignment angle was of order one. If there were an
inflationary epoch between the breaking of the global $U(1)$ symmetry and
$\Tm$ then the majoron field could be correlated over horizon size
regions and could, by chance, be near its
minimum. This would yield a reduction in the misalignment energy density
from the estimate in \eref{rhomis} due to a suppression of $\chi_0$.
The reduction is limited by quantum fluctuations
in the majoron field during the inflationary epoch to a factor of order
$(H_I/\veff)^2$, where $H_I$ is the expansion rate during inflation. One
must also check that isocurvature density perturbations induced by these
fluctuations are not in conflict with observations of the cosmic
microwave background radiation.

One may wonder if the presence of a small explicit symmetry
breaking can cause the initial angle of the majoron field to lie near
the minimum of the majoron potential, \ie\ $\chi_0 \approx 0$.
The same issue is relevant for the question of the formation
of majoron strings and domain walls (see below).
We believe the answer is no, the angular minimum is not strongly
preferred. The essential point is that the radial gradient of
the potential will be much greater than the angular
gradient caused by the explicit symmetry breaking,
see \fref{potential}. Once a VEV has been established, it will roll
radially outward much faster than it can slip sideways.
The one possibility seems to be that the tilt causes the VEV to
be established in a direction aligned with the explicit symmetry
breaking, so that the radial roll of $\phi$ takes one directly
to the global minimum.
To study this, we consider a first order phase transition
which proceeds through the formation of bubbles, which
may form by quantum tunneling or by thermal
fluctuations. We discuss the tunneling case first and comment on thermal
fluctuations later.

The rate of bubble formation depends upon the action to form a
bubble of zero energy, which in turn depends
upon the amount of supercooling and the shape of the potential.
Dimensional analysis suggests that the action
for a critical size bubble is
\elab{acrit}{
A_c \sim R_c^4 \Delta V \sim R_c^3 \sigma \sim
  \frac{\sigma^4}{(\Delta V)^3} ,
}
where $R_c \sim \sigma/\Delta V$ is the radius of a critical size bubble,
$\Delta V$ is the potential difference between the false and
true vacua,
$\sigma \sim v^2/l \sim \Delta V l \sim v (\Delta V)^{1/2}$
is the surface tension on
the bubble, $l \sim v/(\Delta V)^{1/2}$ is the surface thickness, and
$v$ is the magnitude of the VEV to be established after the phase
transition. The formation rate per unit volume scales as
\elab{formrate}{
\Gamma_c \sim R_c^{-4} e^{-A_c}.
}
This rate should be compared to the expansion rate of a horizon volume,
$\sim H^4$. Now, even though
the prefactors may have ratios of large numbers such as $\mpl/T$ this
comparison will be dominated by the exponential. When supercooling first
begins $\Delta V$ will be quite small, the action for bubble formation
will be large, and bubble formation will be slow compared to the expansion
rate. However, as the Universe continues to expand and supercool,
$\Delta V$ will increase and $A_c$ will decrease. Eventually, $A_c$ will
be of order a few hundred and bubble formation will become rapid.
The significant point for us is that by the time this happens the tilt of
the potential will be rather small, $\epsilon \ll \Delta V$. As a result
the action to form a critical bubble with the majoron field at a local
maximum will
be of order $R_c^4 (\epsilon+\Delta V)$ whereas the action to form a bubble at
 the
global minimum of the potential energy is just $R_c^4 \Delta V$.
 Since bubble formation proceeds when
$A_c \sim \hbox{\rm a few hundred}$, we require that
$\epsilon/ \Delta V \sim 10^{-3} $ in order to significantly bias bubble
formation. Since $\Delta V$ is of order $\lambda v^4$,
this requires a very small self coupling for the scalar fields,
which in the context of a renormalizable $U(1)$ local gauge theory is not
natural. We conclude that for the case of quantum tunneling the
majoron field is randomly aligned with respect to the explicit symmetry
breaking.

The thermal fluctuation case is similar. Instead of considering
$\exp(-A_c)$, one considers $\exp(-\beta F_c)$, where $F_c$ is the free
energy of a thermal fluctuation forming a bubble for which
the energy is maximal, $dE/dR = 0$. Which process dominates depends on the
ratio $\beta F_c/A_c$, which in turn depends on the details of the theory.
Generally, the transition for a weakly self-coupled scalar field
should be dominated by thermal fluctuations, whereas for a strongly coupled
field the tunneling process should win out. In either case, the explicit
breaking due to gravity should not be important. Therefore, for a first
order phase transition, even though the
majoron mass may be greater than the expansion rate at the time of
symmetry breaking, there will not be any suppression of the
coherent energy density due to a physical alignment of the majoron field
along a preferred direction, and \eref{omegamis} may be used.

As with the thermal majorons there is the possibility of majoron decay;
however, if we confine ourselves to the case where the coherent energy is
more important than the thermal energy we can show that
majoron decay is not a relevant issue. By using the ratio of coherent to
thermal majoron densities, $\Rn$, given in \eref{Rn}, we can express the
majoron mass as
\elab{mjomegaj}{
\mj = \frac{10^{-8} \omegaj}{\Rn} \GeV.
}
Using this form in \eref{omegamis} gives
\elab{veffomegaj}{
\veff = 10^{8.5} \omegaj^{1/2} \Rn^{1/4} \GeV.
}
To maximize the decay rate we take the neutrino mass to be comparable
to the majoron mass, take $\vnu = \veff$, and find
\elab{taujomegaj}{
\tauj \simgt \frac{10^{17} \Rn^{7/2}}{\omegaj^{5/2}} \sec.
}
Decreasing the lifetime requires either a large value of $\omegaj$ with a
corresponding unacceptable density of decay neutrinos, or requires making
a thermal density of majorons in excess of the coherent density.

\section{Majoron Strings and Walls}

In the models presented here there are several types of topological defects
which may occur. When $U(1)_{B-L}$ breaks, a set of local cosmic strings
forms with string tension $\mu_b \sim \vb^2$. We call these type-$b$
strings since they have a defect in the field $\phib$. Later,
when $T \sim \va$, the second $U(1)$ breaks and type-$a$ global strings
form with $\mu_a \sim \va^2$. At the same time, type-$b$ strings develop
a defect in $\va$ with winding number adjusted to minimize the energy in
the angular gradients of the fields. Since there is explicit symmetry
breaking in our models, at the time $\Tm$ one expects domain walls to form
with surface tension $\sigma \sim \mj \veff^2$. These domain walls may
exist as isolated surfaces or be bounded by strings.
If $\mj >  H(\va)$ these walls will exist from
the moment that majorons exist, but otherwise they form at a later
time.

The fate of the string-wall network depends upon how many walls are
connected to the various types of strings. If either the type-$a$ or the
type-$b$ strings come with a single wall attached, then the system
can dissipate by self intersections~\cite{everett}.
If, however, both types of strings have a number of walls $N_w>1$,
then the strings cannot annihilate and the walls may eventually dominate
the energy density of the universe. The energy density in a single wall
stretching across today's horizon is
\elab{omegawall}{
\omegawall \sim \frac{\sigma H}{H^2 \mpl^2} \sim 10^4
  \frac{\mj \veff^2}{\GeV^3}.
}
A single wall stretching across the horizon would be likely to contribute
to the quadrupole moment of the microwave background, and so it seems
reasonable to require $\omegawall \simlt 10\M5$. The resulting constraint,
$\mj \veff^2 \simlt 10\M9 \GeV^3$ is much
stronger than that which follows from the coherent energy density
(\eref{omegamis}), and somewhat stronger than the constraint from
thermal relics, depending on the magnitude of $\veff$. For example, it is
violated by $\sim 7$ orders of magnitude for the toy model used
in presentation of the thermal relics. If majorons are to make a
significant contribution to $\Omega$, the model must be defined so that
$N_w = 1$ for some set of strings. On the other hand,
it is not difficult to suppress $\mj$
by sufficient powers of $\mpl$ in order that majoron walls pose no
problem.

Constructing models where $N_w = 1$ for either type-$a$ or type-$b$
strings may be accomplished by an appropriate choice of the $B-L$ charges
$\qa$ and $\qb$. The problem is identical to that considered
in reference~\cite{BarrS} for a class of axion models,
so we summarize that analysis. The energy
density in the majoron field around a string is proportional to
\elab{gradtheta}{
\pfrac{\partial \chi}{\partial\theta}^2 \sim (\qb \wa + \qa \wb)^2
    \sim (\nb \wb + \na \wa)^2,
}
where $\wa$ and $\wb$ are the winding numbers for $\phia$ and $\phib$
respectively. Recall that $\na/\nb = \qb/\qa$, with both $\na$ and $\nb$
chosen to be integers with no common factors other than 1.
At the same time, the number of walls is given
by the number of maxima in the gravitationally induced potential
(\eref{operator}),
\elab{numwalls}{
N_W = \nb \wb + \na \wa.
}
Therefore, minimizing the field energy
reduces the number of walls to a minimum as well.

For minimal type-$b$ strings we have $\wb =1$, and $\wa$ is
chosen to minimize the gradient energy; while for type-$a$ strings $\wb = 0$
and $\wa=1$. There are three possibilities for achieving strings with but one
wall. a) If $\nb = 1$, type-$b$ strings have $\wa = 0$ and one wall. b)
Similarly, if $\na = 1$, then type-$a$ strings have a single wall.
c) It is possible for $\na \ne 1$ and $\nb \ne 1$ but to still have
that $\wa$ which minimizes \eref{gradtheta} result in a single wall; \eg\
if $\qa = 2$ and $\qb = 6/5$, then $\nb = 5$, $\na = 3$, and $\wa = 2$.

Even if the network can dissipate, there is an additional constraint that
this happens quickly enough. The
system cannot dissipate until the energy in walls is comparable to the
energy in strings. If the string energy dominates, then small loops of
string that intersect walls result in holes which are subcritical in size;
\ie\ they do not expand to eat up the wall, but rather the hole shrinks
and fills in. In order to get dissipation the critical size hole must be
less than the horizon size. For type-$b$ strings, this occurs when
$(\vb/\va)^2 (H/\mj) > 1$. This may be later than the time when the
majoron mass first becomes effective due to the difference in the
magnitude of the VEVs that control the characteristics of the walls and
the strings. If this time occurs after the time of matter-radiation
decoupling, and $\vb \simgt 10^{16} \GeV$, then the
strings still present at that time
will cause excess fluctuations in the microwave background.

There are additional constraints that arise from the decay products of the
strings and walls. For example, there are the majorons created in the
dissipation of the type-$a$ strings, which strengthen the constraint from
the misalignment energy, \eref{omegamis}. Other examples concern the
type-$b$ strings. After they first form and before they get
dressed up with a majoronic cocoon at $T \sim \va$,
a scaling solution is established and
presumably maintained by the radiation of gravitons. Requiring that the
graviton energy not contribute too much density during nucleosynthesis
provides a constraint on $\vb$,
$\vb \ln(\vb/\va) \simlt 10^{18}\GeV$~\cite{DavisBennett}.
Similarly, if $\na = 1$, type-$b$ strings develop no walls at all for
$\wa = -\nb$. In this case the number density of type-$b$
strings is always determined by a scaling
solution. Requiring that density fluctuations induced by these strings do
not cause excess fluctuations in the microwave background implies
$\vb \simlt 10^{16} \GeV$.

\section{Thermal Majorons and Nucleosynthesis}

If majorons are in thermal equilibrium at $T = 1\MeV$ they will contribute
to the energy density an amount equivalent to 4/7 of a neutrino species.
However, Walker \etal~\cite{WalkerSSOK} argue that based on a determination
of the primordial Helium abundance the excess energy density cannot exceed
the standard cosmology by more than 0.3 species, \ie\ $\deltaN < 0.3$.
As pointed out earlier there are two ways for neutrinos to come into
equilibrium - either through scattering events or through decays and
inverse decays. Bertolini and Steigman~\cite{BertoliniS}
have examined the scattering case
for the simple singlet model discussed in section 2. Using the
cross-section of \eref{sigmanunu} we adapt their result to the
model in section 3 and find
\elab{bbnscat}{
\frac{m_\nu}{\hbox{\rm MeV}} \pfrac{\hbox{\rm GeV}}{\vnu}^2 \pfrac{\qx
 \veff}{\vnu}^2
        < 4.6 \times 10^{-5},
}
where the last factor accounts for the fact that the majoron does not
necessarily couple to the neutrinos. If \eref{bbnscat} is satisfied
the majorons freezeout before $T \sim 100\MeV$ and, as a result, the entropy
released when the pions and muons freeze out dilutes the majorons and their
contribution to the density is less than 0.3 neutrino species. If the
majorons freeze out later they share in the entropy release and the
Helium abundance does not come out right.

The other constraint comes from decays and inverse decays~\cite{babu2}.
As a safe constraint, we require that inverse decays come into equilibrium
after the temperature drops below 5\MeV. If the majorons come into
equilibrium before that, then we can be
sure that they would share in the entropy of the Universe before
nucleosynthesis begins and before the neutrinos themselves
freezeout. If the inverse decays do not become effective until later
they will only share the neutrino energy density. The effects on
Helium production are more difficult to calculate (see, for example,
Dodelson and Turner~\cite{DodelsonT} for a discussion of the subtleties in
phase space distributions during neutrino freezeout).
We therefore take as an approximate condition
\elab{bbndecay}{
\Gamma_i \frac{m_i}{3 \times 5\MeV} \simlt H(5\MeV),
}
where the subscript $_i$ indicates either neutrino or majoron depending
upon which is heavier, $\Gamma_i$ is the decay rate at rest, and the
second factor accounts for the Lorentz dilation of the decay of
relativistic particles. The factor of 3 multiplying the temperature
crudely accounts for the average energy of a massless species.
We ignore cases where two particles are
nearly degenerate in mass, and we take no account of quantum statistics
(Pauli blocking, Bose enhancement) in determining our criteria. Evaluating
\eref{bbndecay} at $\Tid = 5\MeV$ when the number of degrees of freedom is
$g_{eff} = 43/4$,
\elab{bbndecb}{
\frac{\hbox{\rm MeV}}{m_i} \frac{\hbox{\rm sec}}{\tau_i}
  \simgt 1.3 \times 10\M3 .
}

We first consider neutrino decay in the simple singlet model of section 2.
Assuming the neutrino mass is less than 5\MeV, the constraint is
\elab{bbndeca}{
\pfrac{m_\nu}{\hbox{\rm MeV}}^2 \pfrac{\hbox{\rm GeV}}{\vnu}
  \pfrac{\qx \veff}{\vnu} \theta < 5.8 \times 10^{-6},
}
where $\theta = \sqrt{g_{a,ij}^2 + g_{v,ij}^2}$ (see \eref{nudec})
is an off-diagonal coupling strength expected to be much less than
one. If $\theta$ and $\qx \veff/\vnu$ were equal to 1,
the decay constraint is stronger than
the scattering constraint for $v > 14 \GeV$, but for $\theta = 10^{-3}$
they are comparable at $v = 140\GeV$.
For $m_\nu < 30\eV$ and $v > 100\GeV$
neither constraint is restrictive, nor will either constraint become more
strict if we dilute the neutrino-majoron coupling by adding extra scalars.

With explicit symmetry breaking the majoron gets a mass and we must
consider the reactions, $\chi \leftrightarrow \nu\nu$. These reactions are
more interesting than neutrino decay since a) they can proceed through the
diagonal neutrino majoron coupling and b) the majoron mass can be larger
than the neutrino mass and this allows for a faster decay. As an example,
consider our toy model; with $\vnu = \vx = v$, $\qnu = 2$, $\qx = 1/2$,
and $\ggrav = 1$, the constraint is
\elab{bbndecc}{
\pfrac{m_\nu}{\hbox{\rm MeV}}^2 \frac{v}{\hbox{\rm GeV}} \simlt 51.
}
If we confine ourselves to $m_\nu < 30\eV$ then the constraint on $v$ is
quite weak, $v < 10^{10} \GeV$. This is a fictitious constraint - the
majoron mass increases above 10\MeV\ for $v \approx 5 \times 10^4$, and
for masses above this our analysis does not apply.
In \eref{bbndecc} the dependence on $v$ arises due to the $v^{3/2}$
dependence of $\mj$. In other models this dependence will differ, and so
will the constraints, but generally for neutrino masses below 30\eV\
big bang nucleosynthesis does not pose any serious constraints.

\section{Supernova Constraints}

Since the SN 1987A neutrino pulse had its expected duration
$(O(10\sec))$, any scenario where the supernova has its energy drained
more rapidly is ruled out. In this way it is possible~\cite{choi}
to put a constraint on the majoron luminosity.
Assuming the $\rho$ particles are heavy, the largest contribution
to the majoron luminosity
will come from the processes $\nu \nu \rightarrow \chi \chi$, $\nu^\prime
\rightarrow \nu \chi$, and $\nu \rightarrow \nu \chi$, where the last
is matter induced, and is a consequence of the difference in energy between
helicity states~\cite{bere}. Choi and Santamaria\cite{choi} give the
constraints from these processes as
\elab{snscat}{
\nu \nu \rightarrow \chi \chi:
\frac{m_\nu}{\vnu} \frac{\hbox{\rm GeV}}{\vnu} \simlt 2\times 10^{-8},
}
\elab{sndeca}{
\nu^\prime \rightarrow \nu \chi:
\frac{\tau_\nu}{\hbox{\rm sec}} \frac{\hbox{\rm MeV}}{m_\nu}
   \simgt 3\times 10^{-5},
}
\elab{sndecb}{
\nu \rightarrow \nu \chi:
\frac{m_\nu}{\vnu} \simlt 8\times 10^{-7}.
}
The scattering constraint (\eref{snscat}) is slightly stronger than that
from nucleosynthesis (\eref{bbnscat}), whereas the off-diagonal inverse
decay constraint (\eref{sndeca}) is some two orders of magnitude weaker
than the corresponding constraint from nucleosynthesis (\eref{bbndecb}),
and the diagonal argument(\eref{sndecb}) has no analog for the early
Universe. As with the nucleosynthesis constraints, none of the supernova
constraints are relevant if we confine ourselves to
``reasonable'' values, $m_\nu < 30\eV$ and $v > 100\GeV$.
The constraints were based on requiring the majoron luminosity to be less
than $3 \times 10^{53} \erg \sec\M1$ for a 0.85 solar mass core, with a
radius of 10\km, at a temperature of 50\MeV. Although this condition is
somewhat ad-hoc, similar criteria in the study of axion emission have proven
useful as a guide to more detailed numerical
studies~\cite{BurrowsTB}.

In the present context of a massive majoron we need to add inverse majoron
decay as an additional production mechanism. Actually, inverse majoron
decay already occurs in standard majoron models. The decay of a massless
majoron into two positive helicity neutrinos can proceed as the analog to
the helicity flipping neutrino decay discussed by Choi and
Santamaria~\cite{choi}. The
phase space considerations are similar and the matrix element is the same,
so we expect a similar constraint. If the gravitationally induced majoron
mass is larger than the effective neutrino  mass, then the phase
space will be enhanced. We therefore estimate the constraint from inverse
majoron decay by scaling to \eref{sndecb},
\elab{sninvdec}{
\nu \nu \rightarrow \chi:~~{m_\nu\over{\vnu}} \simlt
             8\times 10^{-7} \frac{\mj}{100 \keV},
}
where the neutrino effective mass is of order 100\keV.

The supernova constraints do not apply if the majorons are so strongly
coupled that they are trapped inside a ``majoron-sphere"
which has a black body luminosity
of $<10^{53}\erg/\sec$. Choi and Santamaria estimate that the interaction
$\nu\chi\rightarrow \nu \chi$ will trap the majorons if
\elab{trap}{
 \frac{m_\nu}{\hbox{\rm MeV}} \pfrac{\hbox{\rm GeV}}{\vnu}^2
   \simgt 3\times 10^{-3}.
}
A comparison with the constraints from nucleosynthesis (\eref{bbnscat}) shows
that: if majorons are coupled strongly enough to be trapped, then
they would be brought into equilibrium in the early Universe. Thus,
trapping is not a viable way of evading the supernova
constraint\footnote{For a possible exception to this, see ref
\cite{bmr2}.}.
Rather, the supernova constraints are an extension of the nucleosynthesis
constraint. This is a general result, the thermal conditions in the
supernova are not so different than in the early Universe, but the time
scale required for trapping is shorter than the expansion rate in the
early Universe at comparable temperatures. For example to trap in the core
requires an interaction time less than the light crossing time of
$3 \times 10^{-5} \sec$, whereas the expansion time at
$T = 50\MeV$ is $2 \times 10\M4 \sec$. Actually, there is an even stronger
requirement~\cite{RaffeltS}:
to avoid excess heat transport in the supernova core requires
significantly larger cross-sections than needed just for trapping,
which would lead to even
stronger disagreement with the nucleosynthesis constraint.

The main weakness, apart from numerical uncertainty,
with the constraints summarized above is that they rely
on the standard cold core-bounce model of Type II supernovae being correct,
but that scenario may not be consistent with the
hypothesized majoron parameters.
The crucial point is that if lepton number is strongly violated than the
lepton degeneracy may be erased on infall. The resultant increase in
entropy may cause a ``thermal'' bounce~\cite{FullerMW}, and it is unknown
whether or not such a scenario can produce the observed neutrino pulse from
SN 1987A.

Other constraints may arise if particles produced in the supernova
decay in flight to observable particles while on the way to Earth.
In the majoron/neutrino sector, the most readily observable
particle is the $\antinue$.
Two scenarios have been discussed. a) Heavy neutrinos emitted from the
neutrino sphere can decay producing a $10-20\MeV$ $\antinue$ signal. Such
a signal would have to be temporally separated from the primary neutrino
pulse in order to be readily identifiable~\cite{FriemanHF}.
b) High energy majorons from the core can decay producing a
$\sim 100\MeV$ signal~\cite{DodelsonFT}. Such a signal can be
identified by its spectrum. Indeed, a single high energy event
coincident with the primary neutrino burst would be significant. Since
the neutrino detection cross-sections increase with energy, only a
fraction of the supernova energy must be emitted in high energy particles
to produce a signal. The lack of any high energy events coincident with SN
1987A therefore places a constraint on the production processes for
majorons that decay into $\antinue$'s which is $\sim 1$ order of magnitude
more stringent in the amplitude than the limits in
\eref{snscat}-\ref{e:sndecb}.

\section{Summary}

In this paper we argue that physics at the planck scale may have
interesting consequences for majoron models. Majoron models contain a
global symmetry which, when broken, provides both for neutrino masses and
relatively rapid neutrino decay; however, gravity need not
respect global symmetries. Unless the global symmetry arises as an
automatic consequence of a local symmetry, there is no reason to expect the
symmetry to survive at all in the low energy phenomenology. To this end we
have constructed explicit models where $B-L$ is a spontaneously broken
local gauge symmetry and the majoron arises as the Goldstone boson of a
spontaneously broken `accidental' $U(1)$ symmetry. By varying the $B-L$
charge of a new scalar field in the model, the global symmetry can be
protected against operators with dimension less than some power $d$. The
most interesting cases are those where $d>4$, in which case the explicit
breaking of lepton number is suppressed by $1/\mpl^{(d-4)}$.

The result of explicitly breaking $L$ is that the majoron gets a mass,
which may have interesting consequences. In addition,
by invoking an automatic symmetry
to protect against explicit symmetry breaking the majoron is necessarily
a linear composition of two scalar fields, only one of which couples to
neutrinos. As a result, the majoron coupling to neutrinos may be
suppressed from its value in simpler majoron models.

By appropriate choices of parameters the majorons may be stable, and
thermal relic majorons may play the role of dark matter. Alternatively,
they may decay into light neutrinos with possible consequences for galaxy
formation, or be so light and weakly coupled that they have almost no
observable consequences at all.
If the dimension of the gravitational operators is large enough,
($d \ge 10$), there is the possibility of a coherent energy density
in majorons, similar to that which arises in axion models.

If the majoron mass lies in the MeV region
there will be consequences for big bang nucleosynthesis. For lighter
(keV) majorons, as long as neutrino masses lie below 30\eV\ neither
neutrino annihilation into majorons nor inverse majoron decay
are likely to pose a problem for nucleosynthesis.
Supernova constraints on majoron models are hardly altered by the
majoron's mass, unless it exceeds
a few hundred MeV, which would suppress thermal majoron production in
supernovae. Suppression of the majoron-neutrino
coupling can soften the constraints from supernovae, even for light
majorons.

Finally, there is not complete freedom in choosing the $B-L$ charges for the
scalar fields in the theory. An inappropriate choice implies the formation of
a network of cosmological strings and domain walls that cannot dissipate via
self-intersection. The energy density in such a network would be
prohibitively large, and so such choices are proscribed.

\vskip 0.2in
\noindent{\bf Acknowledgments} \\
DS and KSB are partly supported by DOE grant DE-AC02-78ER05007.
We acknowledge discussions on similar matters with R.N. Mohapatra.
DS thanks S. Petcov, H. Haber and J. Cline for helpful discussions.

\newpage

\fig{eechi}{Leading contribution to $g_{ee\chi}$.}{.1in}

\fig{potential}{Scalar potential showing small explicit breaking.}{.1in}

\end{document}